# Simple analytical approach to the analysis of plasmonic phenomena in dome-shaped protrusions and depressions on metal surfaces

Anton V. Dyshlyuk, Oleg B. Vitrik

*Abstract*—A simple and efficient analytical model is proposed for analyzing plasmonic phenomena arising in nanoscale defects on metal surfaces, specifically low aspect ratio dome-shaped protrusions and depressions. The model enables the calculation, using a straightforward formula, of the polarization charge density on the defect surface and its associated plasmonic near-field, accurately accounting for the defect geometry and incident light wave parameters. It further effectively incorporates the resonant properties of the defects within the electrostatic approximation, making it possible to calculate the correction to the Born approximation for the defect dipole moment. The analytical results obtained under the electrostatic approximation are validated through full-wave numerical simulations utilizing the finite element method. The proposed analytical approach will be valuable for modeling diverse phenomena pertinent to the development of guided-wave plasmonic devices, micro- and nanoscale light manipulation devices, metamaterials, and hybrid optoelectronic circuits, as well as for addressing light scattering problems, especially those involving surface plasmon polariton waves.

*Index Terms*—Localized surface plasmon resonance, surface plasmon polaritons, optical near field, surface defect on metals, Born approximation, light scattering.

## I. INTRODUCTION

COLLECTIVE oscillations of electron density and electromagnetic field that occur at the interface between two materials with real parts of permittivity of different signs (for example, at the metal/dielectric interface), commonly referred to as surface plasmons, exist in two main forms. Firstly, these are surface plasmon polaritons (SPP), which can be considered as special waveguide modes supported by the interface between the two media [1-8]. They can be excited by a phase-matched p-polarized electromagnetic wave incident on the interface. Secondly, there are localized surface plasmons (LSP), which do not propagate being confined to small objects of limited size such as metal nanoparticles or surface nanostructures (SNS). To excite LSP, phase matching is not required and it is sufficient to simply illuminate the nanoparticle or SNS with light of a suitable wavelength. Surface nanostructures supporting LSP can act as mediators, converting light propagating in free space to SPP and vice versa [5-9]. In addition, by varying their geometry, plasmonic SNS can change the parameters of propagating SPP [4]. This opens up broad opportunities for the development of plasmonic devices and metamaterials with unique optical properties and subwavelength circuits for on-chip information processing [16].

At some characteristic frequencies of incident light, local plasmon oscillations are strongly enhanced, which is referred to as local surface plasmon resonance (LSPR) [1-3]. The enhancement of the near electromagnetic field intensity associated with LSPR, brings about a sharp increase in the efficiency of Raman light scattering and photoluminescence of molecules absorbed on the surface of plasmonic nanostructures, which forms the basis of SERS / SEPL methods and similar analytic techniques [9,10]. In addition, the LSPR effect is used in photovoltaic devices to increase the photoelectric conversion efficiency [11, 12], as well as for heat generation at the nanoscale in thermophotovoltaic devices and in photothermal cancer therapy [17].

One of the cheapest, simplest and most efficient methods for producing nanostructures on the surface of plasmonic materials is the so-called laser printing, which consists in the irradiation of the material with ultrashort tightly focused laser pulses. A well-known example of such structures is dome-shaped nanostructures [14,15]. However, despite a wide range of applications of such structures [11, 13-15] there is a lack, to the best of our knowledge, of simple analytical methods to describe plasmonic phenomena in such structures.

Unlike spheres and other separable geometries [1], bumps cannot be solved analytically via Laplace's equation. While numerical approaches are typically used for LSP parameter calculation, we instead apply successive electrostatic approximations to analytically describe: (1) polarization phenomena in low-aspect-ratio dome-shaped protrusions/depressions on metal surfaces, and (2) the associated local fields (interpreted as LSP near-fields). Analytical formulas will be obtained for spatial distribution of such fields and the resonant wavelength of the LSPR for dome-shaped defects. Also, a correction to the Born approximation for the dipole moment of such defects will be calculated. The validity of the obtained analytical relationships will be confirmed by numerical calculations.

Anton V. Dyshlyuk is with the Institute of Automation and Control Processes (IACP) FEB RAS, and Vladivostok State University (VVSU), Vladivostok, Russia (e-mail: anton_dys@iacp.dvo.ru).

Oleg B. Vitrik is with the Institute of Automation and Control Processes (IACP) FEB RAS (e-mail: oleg_vitrik@mail.ru).

## II. RESULTS AND DISCUSSION

Let a plane p-polarized light wave be normally incident from vacuum onto a surface defect in the form of a dome-shaped protrusion (bump) or depression (hole) on the flat surface of an optically thick film of a plasmonic metal with permittivity $\varepsilon_{Me}$ (Fig. 1). We will assume that the profile of the defect is described by G(z), its height $a \ll \lambda$, and its characteristic width b is such that the aspect ratio A=a/b is small. The geometry of the problem under study is two-dimensional, so the defect is infinitely extended along the Y-axis having a dome-shaped cross-section.

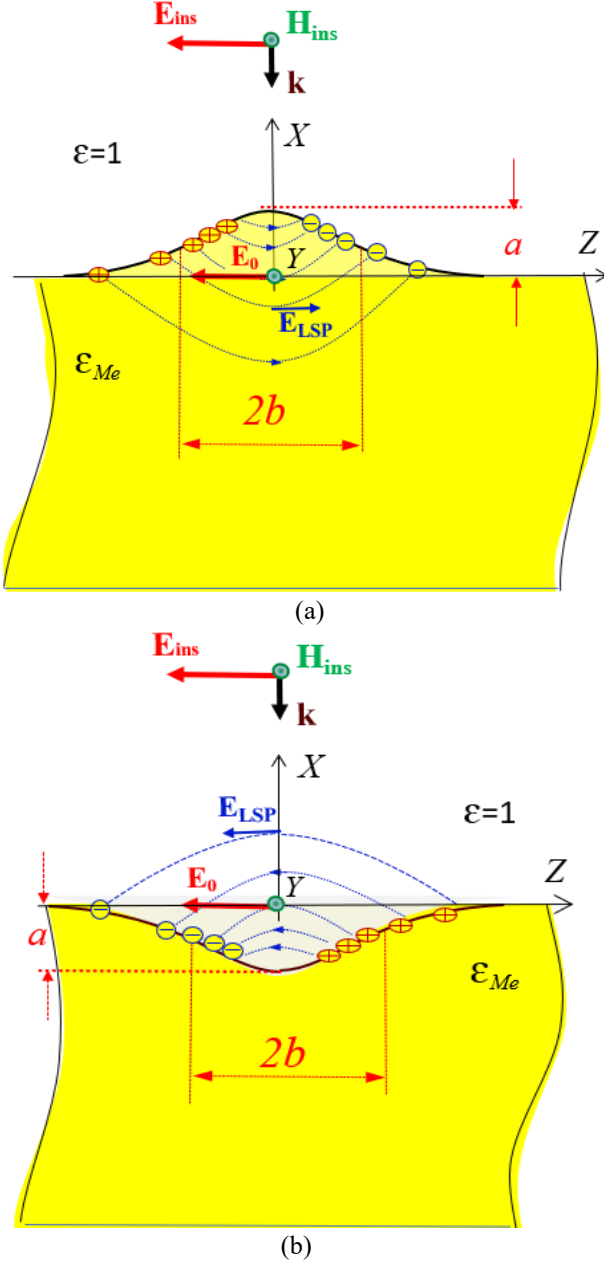

(a)

(b)

**Fig. 1.** The 2D geometry of the problem under study. Figure (a) on the left shows a protrusion, and that on the right (b) shows a depression. The Y-axis in both cases is directed toward the observer.

In the electrostatic approximation, the electric field strength in the vicinity of the defect can be represented as $\boldsymbol{E} = \boldsymbol{E}_0 + \boldsymbol{E}_{LSP}$. In this expression, $\boldsymbol{E}_0$ is the background field strength at the metal surface equal to $\boldsymbol{E}_0 = \boldsymbol{E}_{ins} + \boldsymbol{E}_R$, where $\boldsymbol{E}_{ins}$ and $\boldsymbol{E}_R = r\boldsymbol{E}_{ins}$ are the amplitudes of the incident and reflected waves, respectively, and $r$ is the reflection coefficient, which we will approximately consider equal to that of a flat surface. The field strength $\boldsymbol{E}_{LSP}$ arising due to the polarization charges on the surface of the defect is unknown. This field can also be interpreted as the near field of the defect, i.e. the LSP field. Neglecting the phase shift between the polarization oscillations and $\boldsymbol{E}_0$ the distribution of the field strength $\boldsymbol{E}_{LSP}$ inside the defect can be approximately depicted as shown in Fig. 1.

Let us try to estimate the magnitude of the polarization charge on the surface of the defect. For this, we will use an iterative approach. For simplicity of presentation, we will consider in detail only the case of protrusion (Fig. 2). At the first step, within the 0$^{th}$ approximation, we will neglect the contribution of the field $\boldsymbol{E}_{LSP}$, and the total electric field outside the defect $\boldsymbol{E}^{(0)}$ (the superscript in brackets marks the iteration number) will be considered equal to the background field $\boldsymbol{E}_0$ on the flat metal surface. In a point adjacent to the boundary, the vector $\boldsymbol{E}^{(0)}$ forms angle $\alpha$ with the surface normal (Fig. 2). When penetrating into the medium, the electric field line will be deflected in such a way that the electric field vector $\boldsymbol{E}_s^{(0)}$ in the point directly adjacent to the boundary from inside material, will retain its tangential component, while its normal component will decrease by a factor of $\varepsilon_{Me}$. As a consequence, the normal component of the polarization vector of the medium near the boundary $P_n$ and, accordingly, the surface charge density $\sigma_{surf}$ will be in the first approximation given by $\sigma_{surf}^{(1)} = P_n^{(1)} = \varepsilon_0 \frac{(\varepsilon_{Me}-1)}{\varepsilon_{Me}} E_0 \cos \alpha$.

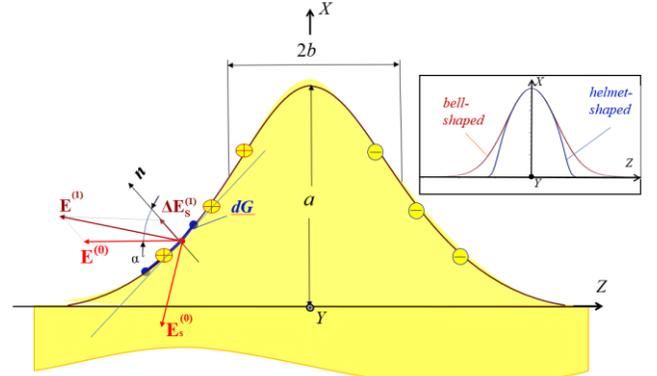

**Fig. 2.** Approximations for the electric field strength near the bump-like defect on the metal surface. The inset illustrates the defect profile.

This rough estimate for $\sigma_{surf}$ can be improved if, at the next iteration step, we take into account the modification of the outside electric field by the field of the surface charge at the adjacent boundary. That is, at this step we set $\boldsymbol{E}^{(1)} = \boldsymbol{E}^{(0)} + \Delta \boldsymbol{E}_s^{(1)}$. The unknown electric field of polarization charges $\Delta \boldsymbol{E}_s^{(1)}$ in the immediate vicinity of the boundary can be approximated by the field of an infinite charged plane, i.e. $\Delta \boldsymbol{E}_s^{(1)} = \boldsymbol{n}_{surf} \sigma_{surf}^{(1)}/2\varepsilon_0$, where $\boldsymbol{n}_{surf}$ is the unit normal to the surface (Fig. 2). Indeed, any finite section of the boundary looks "infinitely" large from a point infinitely close. In ordinary electrostatics, deviations from this approximation arise due to the field of the charges on the remaining surface of the sample, excluding the adjacent elementary section [19]. However, in our "quasi-electrostatic" case, the whole of the conductor is not entirely in the electric field. The electric field of the light wave acts on the conductor locally within the aperture of the light beam, which illuminates only a comparatively small surface area with the defect. Therefore, unlike the situation in electrostatics, the remote parts of the metal surface including those at the opposite side of an optically thick metal layer do not affect the polarization of the defect at all. Note also that an element of the illuminated surface is either almost flat, when it is within the defect with a small aspect ratio (perhaps with the exception of some of its parts with relatively large curvature), or completely flat – outside the defect. This justifies the approximation for the local electric field of the charged surface of the defect - "as the field of a uniformly charged plane". Besides, the validity of such a choice is confirmed by further calculations.

Having an estimate for the modified field $\boldsymbol{E}^{(1)} = \boldsymbol{E}^{(0)} + \Delta\boldsymbol{E}_S^{(1)}$ (Fig. 2), we again consider its deflection at the boundary, which allows us to obtain the next approximation for the surface charge density $\sigma_{surf}^{(2)} = \varepsilon_0 \left( \frac{(\varepsilon_{Me}-1)}{\varepsilon_{Me}} + \frac{(\varepsilon_{Me}-1)^2}{2\varepsilon_{Me}^2} \right) E_0 \cos\alpha$ and, accordingly, the next correction for the field just outside the defect. Proceeding further in this manner we arrive at a geometric progression for the parameter $\sigma_{surf}$, which can readily be summed up with the following result:

$$\sigma_{surf} = 2\varepsilon_0 \frac{\varepsilon_{Me}-1}{\varepsilon_{Me}+1} \frac{dG}{dz} \frac{1}{\sqrt{1+(\frac{dG}{dz})^2}} E_0 \qquad (1)$$

Almost the same result is obtained for the depression, apart from the fact that for a given Z-coordinate, the derivatives $\frac{dG}{dz}$ for the bump and the hole have opposite signs. Accordingly, the surface charge density for the two defects differs by the sign (Fig. 1).

We use two types of profiles as test cases: bell-shaped $G(z) = \pm a \exp(-\frac{z^2}{b^2})$ and helmet-shaped $G(z) = \pm a \exp(-\frac{z^2}{b^2} - \frac{z^8}{b^8})$. Note that with the same aspect ratio $A=a/b$, the minimum radius of curvature of the helmet-shaped dome (equal to 0.127 $a/A^2$ near the base) is four times smaller than the minimum radius of curvature of the bell-shaped dome (0.5 $a/A^2$ near the top). As can be seen, the surface of the "helmet" is more curved and, in this sense, differs from a plane significantly more than the surface of the "bell". Therefore, the "helmet" is a more difficult test case for verifying the validity of relation (1).

The results of charge density calculation for the bumps and holes with the above-mentioned profiles on the surface of gold, using expression (1), at λ= 0.7 μm, are presented in Fig. 3. To verify the analytic results, we plot in the same figure the results of numerical calculation of the surface charge density due to the polarization of bell-shaped (Fig. 3(a, b)) and helmet-shaped (Fig. 3(c, d)) bumps and holes by a normally incident plane wave. The numerical results were obtained by solving the monochromatic 2D Helmholtz equation with the finite element method in the COMSOL Multiphysics software. The size of the calculation domain and the mesh resolution were chosen so that they did not have a significant effect on the numerical results. Perfectly matched layers (PML) were used to absorb scattered radiation at the outer boundaries of the calculation domain.

As can be seen from the figure, good agreement between the analytical and numerical dependences is observed for both the bell-like and helmet-like profiles, as long as the aspect ratio $A$ does not exceed 0.2. This confirms the validity of the "flat" approximation made above for the field of polarization charges at the surface of the defects.

At larger values of $A$, significant discrepancy is observed between the analytical and numerical results primarily near the base of the helmet-like bumps and holes, where, as noted above, the surface curvature is the largest (Fig. 3c,d). Secondly, when A>0.2, the 'opposite sign rule' for the densities of polarization charges in bumps and holes no longer holds. Notably, for the protrusion the analytical and numerical results remain consistent at least up to A~0.5. However, for the depression, the analytical approach yields an underestimated result as early as at A ≳ 0.2. This discrepancy may be attributed to an additional cavity resonance in high-aspect-ratio holes in gold. This leads to an enhanced electromagnetic field in the hole and, consequently, to a further increase in polarization effects compared to the analytical model.

In Fig. 3, panels (e) and (f) again show the results of numerical calculations for the surface charge density of bell-shaped protrusions and depressions at λ=0.7 μm. However, unlike the results presented in panels (a) and (b), the aspect ratio A is fixed at 0.2, while the absolute dimensions of the defects vary. It is evident that good agreement between numerical and analytical data is observed as long as a≲5 nm (and, consequently, b≲25 nm). The increasing discrepancy at larger sizes is likely related to the breaking down of the electrostatic approximation in the analytical model.

Using Coulomb's law and expression (1), one can calculate the two Cartesian components of the field strength $\Delta\boldsymbol{E}_{LSP}$ created by polarization charges at an arbitrary point $(z_0, x_0)$:

$$E_{LSPz}^{(\pm)}(z_0, x_0) = \mp E_0 \frac{1}{\pi} \frac{\varepsilon_{Me}-1}{\varepsilon_{Me}+1} \int_{-\infty}^{\infty} \frac{z_0-z}{(z_0-z)^2+(x_0-G(z))^2} \frac{dG}{dz} dz \qquad (2a)$$

$$E_{LSPx}^{(\pm)}(z_0, x_0) = -E_0 \frac{1}{\pi} \frac{\varepsilon_{Me}-1}{\varepsilon_{Me}+1} \int_{-\infty}^{\infty} \frac{(x_0-G(z))}{(z_0-z)^2+(x_0-G(z))^2} \frac{dG}{dz} dz \qquad (2b)$$

In these expressions and below, the superscript "+" corresponds to the protrusion, and "-" to the depression.

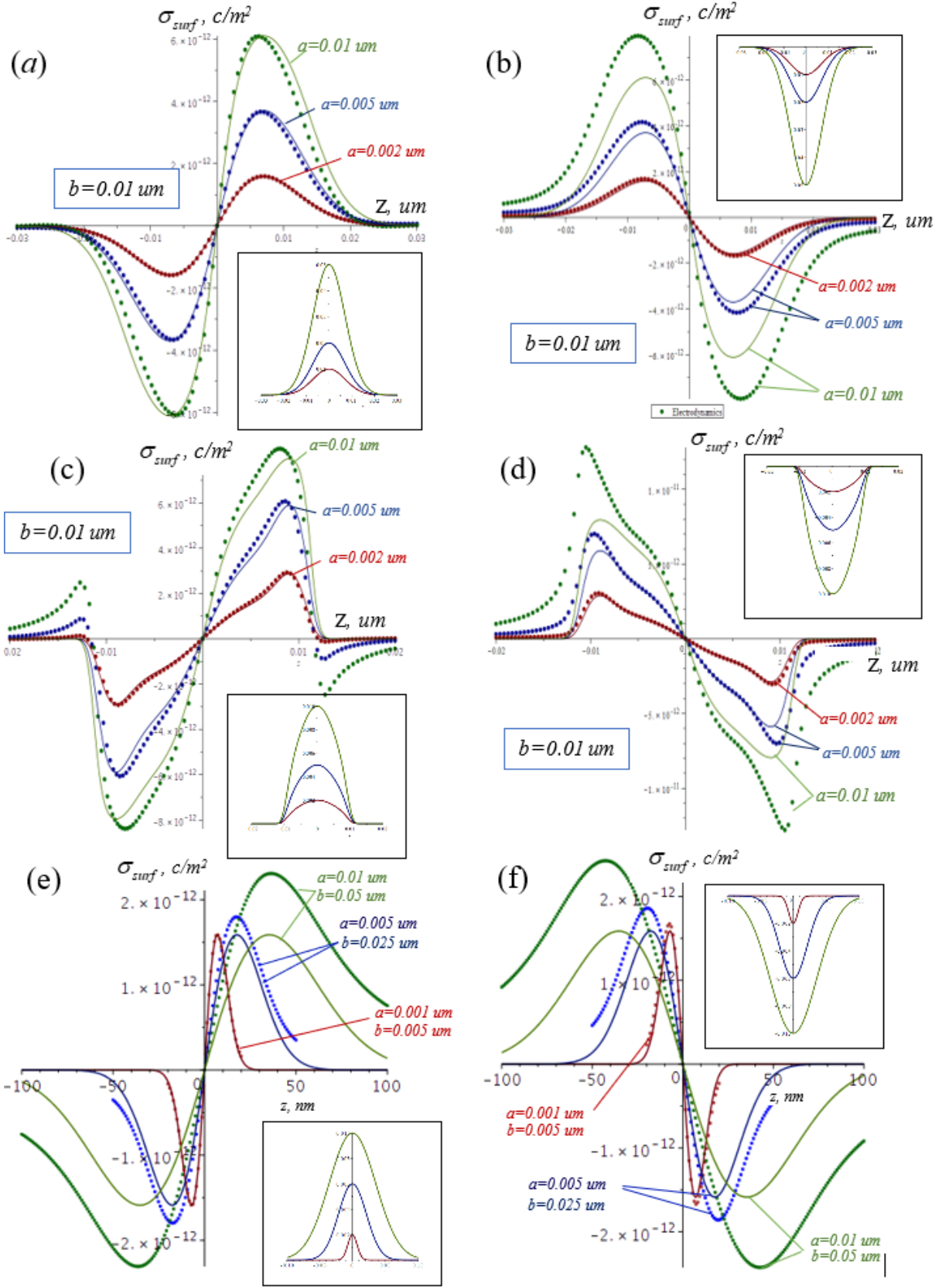

**Fig.3.** Results of calculating the surface charge density arising due to the polarization of the defect: a, b – respectively, a bell-shaped bump and hole ($G(z) = \pm a\, exp(-\frac{z^2}{b^2})$) with a width of b = 0.01 µm ; c, d – respectively, a helmet-shaped bump and hole ($G(z) = \pm a\, exp(-\frac{z^2}{b^2} - \frac{z^8}{b^8})$), b=0.01 µm; e, f – respectively, bell-shaped bump and hole with fixed aspect ratio A = 0.2 , but with different values of b. The profiles of the inhomogeneities are illustrated in the insets to the figures in the corresponding colors. In all figures, the wavelength of the incident light is 0.7 µm.

The distributions of the vector field $\mathbf{E}_{LSP}$ of the polarization charges on the surface of bell-shaped defects calculated with expressions (2) are shown by arrows in Fig. 4(a) in the "analytical" column (under the "+" sign for the bump and "-" for the hole). In Fig. 4(b), in the "analytical" column, also shown by arrows is the distribution of the total field $E = E_0 + E_{LSP}$. The distribution of the square of the amplitude of the corresponding fields in these figures is shown with a color map.

From the figures presented, one can notice the interference enhancement of the total field $E$ inside and in the immediate vicinity of the depression and its weakening inside and near the protrusion. This effect is easily explained by examining Fig. 1, from which it is evident that the near "quasi-electrostatic" field of surface charges $E_{LSP}$ is such that, the z-component of the field $E_{LSP}$ always adds up with the corresponding component of the field $E_0$, enhancing the total field $E$ inside the depression. In the bump, on the contrary, the z-components of the fields $E_{LSP}$ and $E_0$ interfere destructively suppressing the total field.

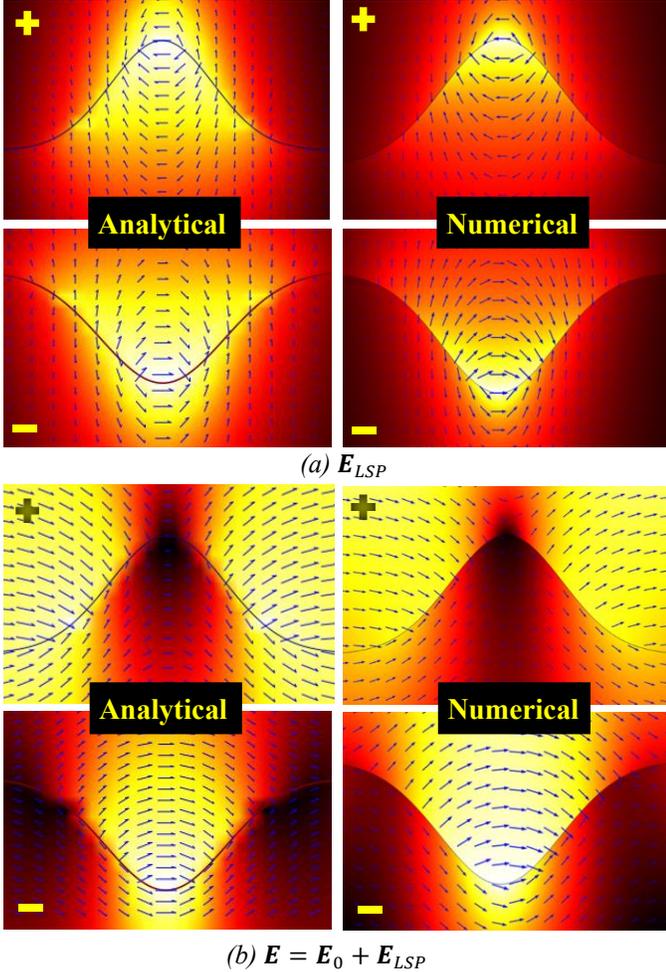

**Fig. 4.** Distribution of the electric field in the vicinity of a bell-shaped defect of width b = 10 nm and height a = 5 nm. (a) Defect's own electric field (due to the polarization charges) calculated analytically (on the left) and numerically (on the right), (b) the total field (incident wave + defect's own field) calculated analytically (on the left) and numerically (on the right). In all figures, the designation "+" refers to the protrusion and "-" to the depression. The wavelength of light is 0.7 microns.

Fig. 4 also shows, in the column "Numerical", numerically calculated fields $E_{LSP}$ and $E$ near bell-shaped defects ("+" for protrusions and "-" for depressions) obtained by the finite element method in the frequency domain using COMSOL Multiphysics software. A good agreement between analytical and numerical results is evident. In particular, both demonstrate similar behavior of the total field vector near the defect: in the metal, the vectors are oriented so as to almost follow the bends of the surface, and outside – they are nearly normal to the surface. Additionally, the numerical results confirm the effect of the weakening of the total field inside the bump and its enhancement in the hole.

Since the color map in Fig. 4 provides rather a qualitative understanding of the electric field amplitude distribution, we also plot in Fig. 5 numerically (dots) and analytically (solid curves) calculated field slices for both Cartesian components of $E_{LSP}^{(+)}$ (z) at different values of x (as measured from the flat metal surface, see the insets). The calculations are carried out for the bell-shaped protrusions with b = 10 nm, λ = 700 nm, and a = 2 nm (a) and a = 5 nm (b). It is evident that the analytical and numerical dependences exhibit almost the same discontinuity at the surface of the defect, and agree well with each other both inside and outside it.

Similarly to what was observed for the charge density $\sigma_{surf}$, the agreement between the analytical and numerical results for $E_{LSP}$ is observed for the protrusion in a much wider range of $A$ than for the depression. This is illustrated in Fig. 6, which shows the results of numerical (dots) and analytical (solid curves) calculations of the amplitude of $E_{LSPz}$ at the center (z = 0, x = a/2) of the bell-shaped protrusion (marker "+", red) and depression (marker "-", blue) vs. their height / depth. The width of the defect is fixed at b = 10 nm and the wavelength of the incident light is λ=0.7 μm. As one can see, while for the protrusion the numerical and analytical results are in good agreement for $A \lesssim 2$, in the case of the depression they agree well only for $A \lesssim 0.2$, with the analytical calculation underestimating the absolute value of $E_{LSPz}$ at larger aspect ratios. Same as above, we attribute this effect to the geometric resonant properties of the holes.

Fig. 7 presents the results of analytical calculations of $\left|E_{LSPz}^{(+)}\right|$ at the center (at z=0, x=a/2) of the bell-shaped protrusion with $b$ = 10 nm vs. incident light wavelength, carried out using expression (2) for two heights of the defect: $a$ = 2 and 5 nm, chosen so that the aspect ratio does not exceed 0.5. As clear from the figure, the dependences have a resonant character, which can be attributed to the factor $(\varepsilon_{Me} - 1)/(\varepsilon_{Me} + 1)$ in expression (2) and which is identical the corresponding factor describing the dipole resonance of a thin cylindrical wire [1]. Same as for the wire, the resonance of the dome-shaped under study would be sharp in the absence of losses (i.e. $Im(\varepsilon_{Me}) = 0$) and would be observed at $Re(\varepsilon_{Me}) = -1$, i.e. in the ultraviolet region for gold. Ohmic losses in real gold lead to a significant red shift and broadening of the resonance as observed in Fig. 7.

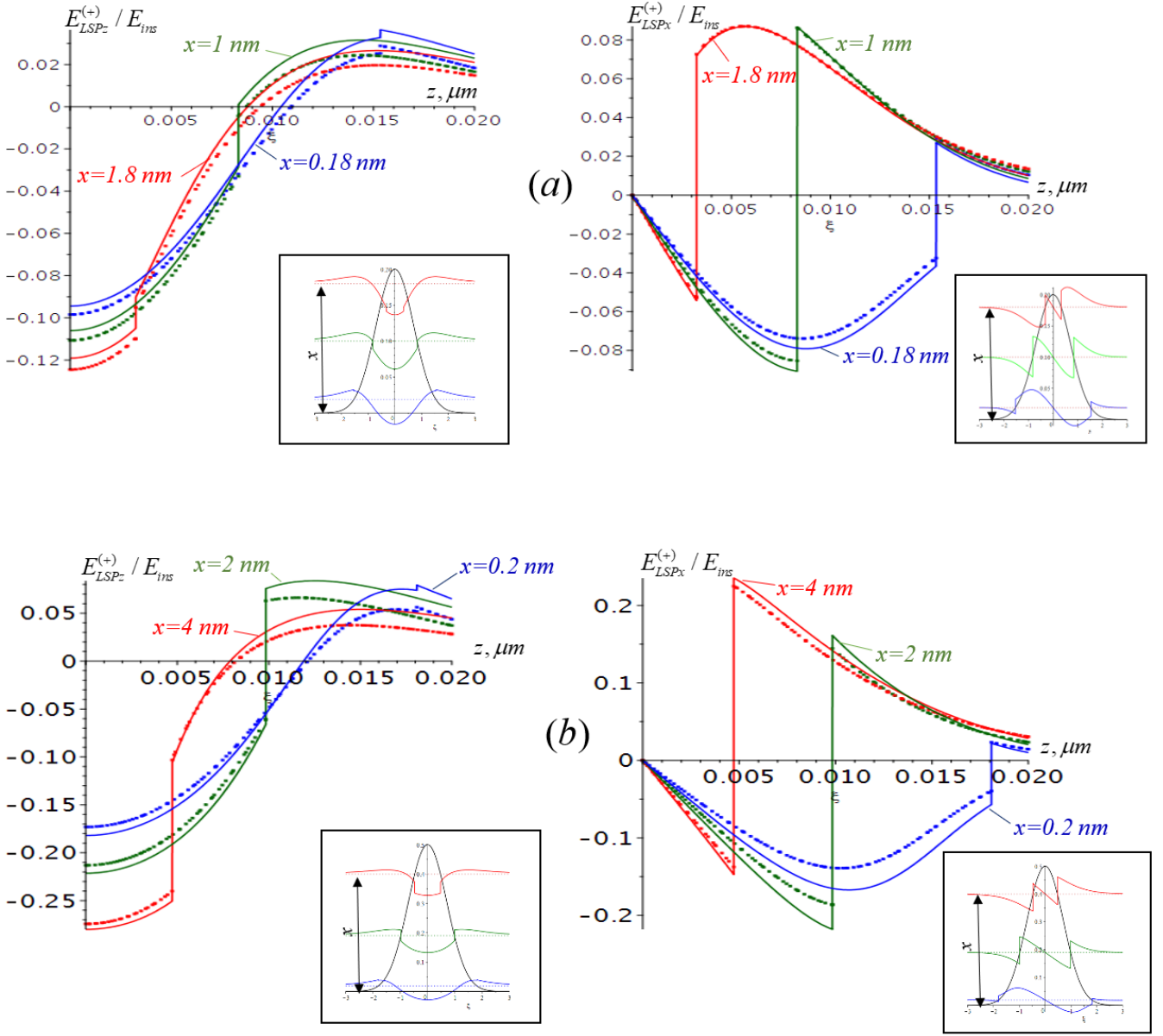

**Fig. 5.** Z-slices of the z-(on the left) and x-components (on the right) of the electric field of the polarization charges $E_{LSP}^{(+)}(x, z)$ at various values of x (as shown the insets, which also reproduce the field slices (against the background of the defect profile) vertically offset by the corresponding value of x) for the bell-shaped protrusion with b = 10 nm and a = 2 nm (a) and a = 5 nm (b). Due to symmetry, only half of the field slice needs to be shown (e.g., for z>0), as is done for clarity.

Also shown in Fig.7 are the results of corresponding numerical calculations of $|E_{LSPz}^{(+)}|$. It is evident that numerical and analytical results are in good agreement.

As is known, defects on the metal surface can be treated as sources of induced currents giving rise to light scattering and excitation of surface plasmon polaritons, with the corresponding current density given by [20]: $J_S = -ik\varepsilon_0(\varepsilon(z,x) - \varepsilon_S(x))\frac{E}{\rho_V}$, where $\rho_V = \sqrt{\mu_0/\varepsilon_0}$ is the wave impedance of the vacuum, functions $\varepsilon(x, z)$ and $\varepsilon_s(x)$ describe the spatial distribution of the permittivity of the metal with perturbed and unperturbed (i.e. flat) boundaries, respectively. It is easy to see that with this approach, the induced current exists only inside the defect and is located above the surface for the the protrusion, and below it for the depression. The current density, therefore, can be rewritten as $J_S^{(\pm)} = \pm i\omega P$, where $P = -\varepsilon_0(\varepsilon_{Me} - 1)E$. For a subwavelength defect, it is not the current density itself that is important, but the total current, and corresponding total dipole moment. Before calculating the latter, let us briefly revisit Fig. 5. In this figure, especially in the insets, it is clearly seen that the distribution of the z-component of $E_{LSP}$ inside the defect is an even function of z, while that of the x-component is an odd function. This means that only the z-component of $E_{LSP}$ will contribute to the total dipole moment $p$, which, therefore, will have only the z-component. In this case, the problem of calculating p (since our geometry is 2D, $p$ should be given per unit length $l$ in the transverse direction) is reduced to calculating:

$$pl^{-1} = p_B l^{-1} + p_{LSPz}^{(\pm)} l^{-1}, \qquad (3)$$

where $p_B l^{-1} = -\varepsilon_0(\varepsilon_{Me} - 1)E_0 S$, S is the cross-section area of the defect, $p^{(\pm)}_{LSPz} l^{-1} = \mp \varepsilon_0 (\varepsilon_{Me} - 1) \iint_{(S)} E_{LSPz}\, dzdx$, with (S) below the double integral sign meaning that the integration is carried out over the defect cross-section.

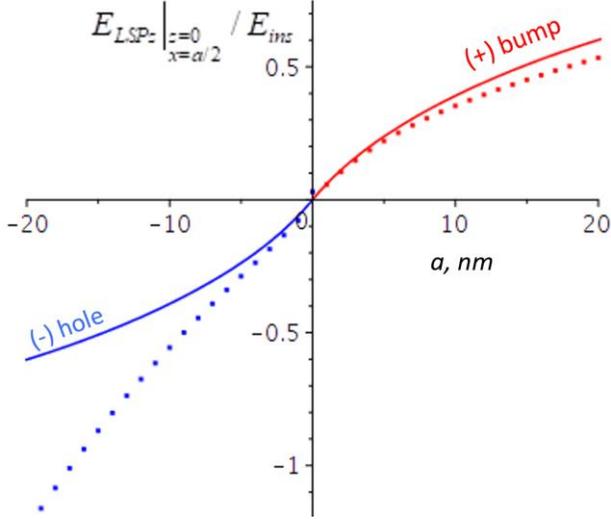

**Fig. 6.** Dependence of the amplitude of $E_{LSPz}$ at the center (at $z = 0$, $x = a/2$) of the bell-shaped protrusion (a>0) and depression (a<0) on their height/depth. The width of the defect is b = 10 nm and the wavelength of the incident light is λ=0.7 μm.

The first term in expression (3) corresponds to the Born approximation [18] for the dipole moment, the second is the correction to this approximation due to polarization phenomena. Assuming a bell-shaped profile for the defect and expanding expression (2a) in a series with respect to $a$ up to the first order, after some transformations we obtain the following expression for the polarization correction:

$$p^{(\pm)}_{LSPz} l^{-1} = \mp \varepsilon_0 \frac{(\varepsilon_{Me}-1)^2}{\varepsilon_{Me}+1} a^2\, Im\left( Q(\sqrt{\ln 2} - i\frac{a}{2b}) \right) E_0, \quad (4)$$

where Q(x) is defined as: $Q(x) = e^{-x^2 + 1/2}(\mathrm{erf}(ix) - 1)$.

Solid curves in Fig. 8 show analytically calculated dependences of the total dipole moment of the bell-shaped defect on its height / depth using expressions (3) and (4). In the calculations, it is assumed that the width of the defect is b=10 nm, the wavelength of the incident light is λ= 0.7 μm. Curve 1 in Fig. 8 was obtained in the Born approximation, taking into account only the first term in expression (3). Curves 2 and 3 were obtained, respectively, for the protrusion and depression, taking into account the polarization corrections as given by expression (4).

In Fig. 8, the dots show the results of numerical calculations of the total dipole moment of the bell-shaped bump (marker "+") and hole (marker "-") depending on the height / depth of the defect, at b=10 nm, λ=0.7 μm. For the bump, these results clearly agree well with the analytical calculations. As for the depression, as expected, the analytical approach underestimates the polarization correction at $a \gtrsim$ 2nm (A ≳ 0.2).

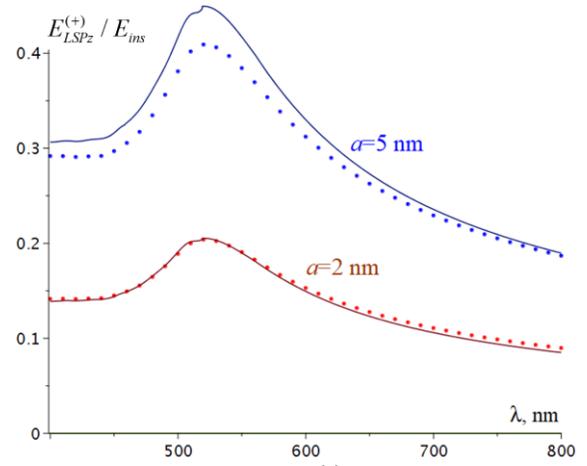

**Fig. 7** Dependence of the amplitude of $E^{(+)}_{LSPz}$ at the center (z=0, x=a/2) of the bell-shaped protrusion, on the incident light wavelength. The width of the defect is b = 10 nm. Solid lines are the result of analytical calculations, while dots indicate numerical results.

## V. CONCLUSION

Thus, the proposed analytical model provides a good overall description of the dependence of both the surface polarization charge density and the local LSP field on the parameters of the incident electromagnetic wave and the geometry of low-aspect-ratio subwavelength dome-shaped defects. It also effectively accounts for the resonant properties of such defects within the electrostatic approximation, enabling the calculation of the correction to the Born approximation for the defect dipole moment. We anticipate that this approach will prove beneficial for modeling various phenomena involved in the development of plasmonic devices, micro-nanoscale light manipulation guided-wave devices, metamaterials, and hybrid optoelectronic circuits, as well as for solving light scattering problems, including those dealing with SPP waves.

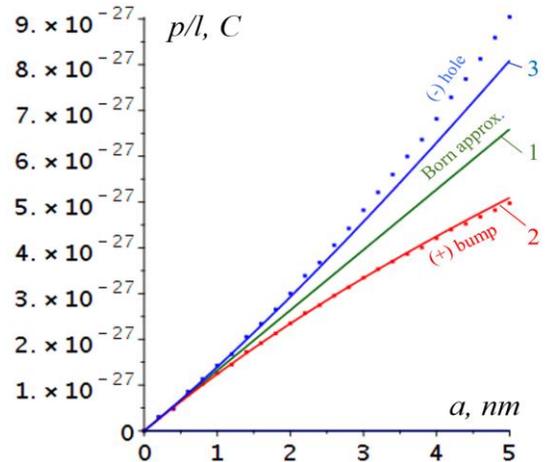

Fig. 8 Dependence of the total dipole moment of a bell-shaped bump and hole per unit length in the transverse direction on their height / depth. The width of the deffect is b = 10 nm. Incident light wavelength is λ= 0.7 μm. Curve (1) is an analytical result obtained in the Born approximation, (2) and (3) are the analytical results for the protrusion (2) and the depression (3), including polarization correction. The dots of the corresponding color show the results of numerical simulation.